\documentclass[aps,prl,twocolumn,nopacs,floats,superscriptaddress]{revtex4-1}
\usepackage{graphicx}
\usepackage{dcolumn}
\usepackage{bm}
\usepackage{amsmath}
\usepackage{color}
\usepackage{amssymb}
\usepackage{ulem}
\usepackage{xcolor}
\usepackage{subfigure}
\usepackage{ulem}

\definecolor{cardinal}{rgb}{0.6,0,0}
\definecolor{darkgreen}{rgb}{0,0.4,0}
\definecolor{golden}{rgb}{0.92, 0.7, 0}
\definecolor{midnight}{rgb}{0, 0, 0.5}
\definecolor{darkblue}{rgb}{0, 0, 0.7}

\def\he4{$^4$He}
\def\hel3{$^3$He}
\def\Am3{\AA$^{-3}$}
\def\beq{\begin{equation}}
\def\eeq{\end{equation}}

\newcommand{\be}{\begin{equation}}
\newcommand{\ee}{\end{equation}}
\newcommand{\bea}{\begin{eqnarray}}
\newcommand{\eea}{\end{eqnarray}}
\newcommand{\bse}{\begin{subequations}}
\newcommand{\ese}{\end{subequations}}
\def\rf#1{(\ref{#1})}

\begin{document}

\author{Anatoly Kuklov}
\affiliation{Department of Physics \& Astronomy, College of Staten Island and the Graduate Center of
CUNY, Staten Island, NY 10314}

\author{Leo Radzihovsky}
\affiliation{ Department of Physics and Center for Theory of Quantum Matter, University of Colorado, Boulder, CO 80309}

\author{Boris Svistunov}
\affiliation{Department of Physics, University of Massachusetts, Amherst, MA 01003, USA}

\title{Field Theory of Borromean Super-counterfluids}
\begin{abstract}
We introduce a class of dynamical field theories for $N$-component  “Borromean" ($N\geq 3$) super-counterfluid order,  naturally formulated in terms of inter-species bosonic fields $\psi_{\alpha\beta}$. Their condensation breaks the normal-state [U(1)]$^N$ symmetry down to its diagonal U(1) subgroup, thereby encoding the arrest of the net superflow. This approach broadens our understanding of dynamical properties of super-counterfluids, at low energies capturing its universal properties, phase transition, counterflow vortices, and many of its other properties. Such super-counterfluid strikingly exhibits $N$ distinct flavors of energetically stable elementary 
 vortex solutions, despite $\mathbb{Z}^{N-1}$ homotopy group of its $N\! -\! 1$ independent Goldstone modes, with $N\! -\! 1$ topologically distinct elementary vortex types, obeying modular arithmetic. The model leads to Borromean hydrodynamics as a low-energy theory, reveals counteflow AC Josephson effect, and generically predicts a first-order character of the phase transitions into Borromean super-counterfluid state in dimensions greater than two.  
\end{abstract}

\maketitle

{\it Introduction.}
Super-counterfluidity \footnote{Also known as counterflow superfluidity.} is the phenomenon of inter-flavor superfluid-like off-diagonal order in a multicomponent (either bosonic or fermionic) system under the condition of arrested net superflow. At zero temperature, the super-counterfluid (SCF) ground state is driven by Mott physics, requiring exact commensurability between total particle density and the underlying lattice \cite{KS2003}.  A related counter-flow states also appear within a Landau-level \cite{BalentsLR96} and other constrained dynamical systems as dipole Bose-condensates \cite{Bakr20,Du22,Lake22}. At finite temperature SCF is driven by proliferation of a subset of multicomponent composite vortices \cite{Babaev2002,KPS2004}. 

Another type of a SCF state emerges when multiple fields of a multicomponent superconductor couple to a single U(1) gauge field, thereby, through the Meissner effect suppressing the net matter flow in the bulk of the system, but allowing for the neutral SCF modes \cite{Babaev2004} (see also Ref.~\cite{sbpbook}).

A particularly interesting form of SCF---the so-called Borromean SCF (SCF$_B$)---arises for $N \geq 3$ components \cite{sbpbook,Blomquist2021,BS2024,Golic2025}. The qualitatively distinct physics of SCF$_B$ is characterized by the presence of $N$ distinct flavors of energetically stable vortex solutions, with modular arithmetic of their topological charges, and persistent counterflow states. This is despite only $N-1$ of them are topologically independent, as dictated by the homotopy of [U(1)]$^N$/U(1) being $\mathbb{Z}^{N-1}$, with SCF characterized by $N\! -\! 1$ independent Goldstone modes.

SCF$_B$ is also a natural mother state for a novel time-reversal-breaking ``Borromean insulator," driven by  a frustrating inter-component Josephson coupling \cite{Bojesen2013,Bojesen2014, BS2024}. Recent experiments  on  Ba$_{1-x}$K$_x$Fe$_2$As$_2$
were naturally interpreted in terms of this phase at finite-temperature \cite{Grinenko2021,Shipulin2023}.

It was only recently that the effective hydrodynamic (phase-density) description of SCF$_B$ state was developed \cite{BS2024,Golic2025}, capturing the essence of the long-scale Borromean counterflow and demonstrating its qualitative distinction from mere particle pairwise counter-propagation at the microscopic level. The unifying principle behind Borromean hydrodynamics is the {\it local} gauge redundancy---vanishing stiffness---of the phase common to all $N$ components, Eq.~(\ref{compact_guage}) (dubbed ``compact-gauge invariance" \cite{BS2024}). This contrasts standard local  U(1) gauge redundancy encoded through a common gauge field, as in, e.g., multicomponent superconductors.  By Noether's theorem, this leads to vanishing of the total current, thus revealing that SCF$_B$ is an insulator with respect to the total charge (not to be confused with previously-mentioned ``Borromean insulator"). However, dynamic (Gross-Pitaevskii type) field theory, providing a more complete description of the Borromean SCF 
has been missing, and it is our goal here to develop a minimal field theory of this phenomenon. 

In this Letter, we show that the 
Borromean SCF state---characterized by the broken symmetry [U(1)]$^N$/U(1)---can be naturally described by an $N(N\! -\! 1)/2$ component 
composite tensor field $\psi_{\alpha \beta}$, transforming 
identically to the direct product of  U(1) fields.  Treating $\psi_{\alpha\beta}$ as a fundamental complex field,  we formulate  dynamic field theory for $\psi_{\alpha\beta}$, that, at low energies, captures all universal properties of the Borromean super-counterflow fluid. The model leads to Borromean hydrodynamics of the ordered state, reveals counteflow AC Josephson effect, $N$ counterflow vortices, and generically predicts a first-order character of the phase transitions into SCF$_B$ state in dimensions greater than two.

{\it SCF$_B$ model.} We now develop a dynamical field theory that exhibits  SCF$_B$  described by the $N(N-1)/2$-component Hermitian order parameter
\be
\psi_{\beta \alpha}\,  \equiv \psi_{\alpha \beta}^* \;,
\label{conv}
\ee  
a tensor field with $\alpha,\beta =1, \, 2,\,  \ldots , \, N$, and $\psi_{\alpha \alpha} = 0$.  Such order parameter naturally transforms under the factor group [U(1)]$^N$/U(1), with a one-dimensional representation $\hat{U}_\alpha(\varphi) \psi_{\alpha \beta}= e^{i \varphi }\psi_{\alpha \beta} $, $\hat{U}_\beta(\varphi) \psi_{\alpha \beta}= e^{-i \varphi }\psi_{\alpha \beta} $, and $\hat{U}_\alpha(\varphi) \psi_{\mu \nu}=\psi_{\mu \nu}$,  $\mu\neq \alpha$, $\nu\neq \alpha$.  As required, $\psi_{\alpha\beta}$ is a singlet under the diagonal subgroup, $\left[\prod^N_{\gamma=1} \hat{U}_\gamma (\varphi)\right]\psi_{\alpha\beta}=\psi_{\alpha\beta}$.

In the Hamiltonian formalism, the $N(N-1)/2$ fields $\psi_{\alpha\beta}$ ($\beta > \alpha$) are complex canonical variables, with  $\psi_{\beta \alpha}$ the corresponding canonically (also complex-) conjugate fields (cf.~Ref.~\cite{sbpbook}). We propose a generic SCF$_B$ Hamiltonian density that consists of two parts,
\be
{\cal H}\, =\, {\cal H}_0 \, +\, {\cal H}_\text{couple} \, ,\label{H}
\ee
with ${\cal H}_0$ a sum of independent Ginzburg-Landau/Gross-Pitaevskii type Hamiltonians for each of the $N(N-1)/2$ canonical fields $\psi_{\alpha\beta}$ ($\beta > \alpha$).  To lowest order ${\cal H}_0$ is given by
\be
{\cal H}_0 \, =\, {1\over 2}\sum_{\alpha \neq \beta} 
\left( {\Lambda_{\alpha \beta} \over 2} |\nabla \psi_{\alpha \beta}|^2 \, +\, {U_{\alpha \beta} \over 2} |\psi_{\alpha \beta}|^4 \right) ,
\ee
with  $\Lambda_{\alpha \beta}=\Lambda_{ \beta \alpha}$ and $U_{\alpha \beta} = U_{ \beta \alpha}$ being positive definite couplings; the global factor $1/2$ compensates for the double counting of each canonical degree of freedom. We note that ${\cal H}^{(0)}$ has an enlarged symmetry of [U(1)]$^{N(N-1)/2}$, with each component of the tensor field $\psi_{\alpha \beta}$ transforming under independent U(1): $\hat{U}_{\alpha\beta}\psi_{\alpha \beta}= e^{i \varphi_{\alpha\beta}}\psi_{\alpha \beta} $.

${\cal H}_\text{couple}$ plays the key role in reducing this enlarged symmetry down to the physical  [U(1)]$^N$/U(1) symmetry 
through a cubic phase-locking \footnote{Phase independent density-density coupling plays little role. Higher-order phase-sensitive terms are subdominant (though may become important for large coupling): Without a qualitative difference, it is possible to add any number of order-$s$ ($s > 3$) terms such that each of them has the cyclic form $\psi_{\alpha_1 \alpha_2} \psi_{\alpha_2 \alpha_3}\ldots \psi_{\alpha_s \alpha_1}$ guaranteeing invariance under the Borromean symmetry group.},
\be
{\cal H}_\text{couple}\, = \, - \! \sum_{\{ \alpha \beta \gamma\}} \!  g_{\alpha \beta \gamma} (\psi_{\alpha \beta} \psi_{\beta \gamma} \psi_{\gamma \alpha}
+\psi^*_{\alpha \beta} \psi^*_{\beta \gamma} \psi^*_{\gamma \alpha})
\label{V_terms}
\ee
that we refer to as {\it Borromean coupling}.
Here the sum is over all distinct triplets $\{ \alpha \beta \gamma\}$ with $g_{\alpha \beta \gamma}$ the corresponding positive coupling constants. 

Consistent with SCF$_B$'s $N \! -\! 1$ broken continuous symmetries, the ground state of Hamiltonian (\ref{H})--(\ref{V_terms}) supports $N\! -\! 1$ Goldstone modes, with the rest of the modes gapped out by  ${\cal H}_\text{couple}$.  To see this we note that  Borromean coupling can be conveniently interpreted as enforcing (at low energy) the vanishing (mod $2\pi$) of the ``flux" $\phi_{\alpha \beta} + \phi_{\beta \gamma} + \phi_{\gamma \alpha}$ through all $\alpha\beta\gamma$ ``plaquettes," thereby constraining the bond phases of $\psi_{\alpha\beta}\sim e^{i\phi_{\alpha\beta}}$ to be expressible in terms of the $N\! -\! 1$ phases $\theta_\alpha$ [diagonal subgroup remains unbroken, with $H$ in (\ref{H}) thereby independent of the overall phase] of the associated “plaquette" vertices, 
\be
\forall \, \{ \alpha \beta \}:  \quad \phi_{\alpha \beta} \, =\, \theta_\alpha - \theta_\beta \, .
\label{theta}
\ee

This symmetry and Goldstone mode reduction to $N-1$ can also be explicitly demonstrated  by the following parameterization of the phases  of the fields $\psi_{\alpha \beta}$. For each pair $(\alpha, \beta)$ such that, say, $\alpha \neq 1$, $\beta \neq 1$, we can parameterize $\phi_{\alpha \beta}$  as  $\phi_{\alpha \beta} = \phi_{\alpha 1} + \phi_{1 \beta} + \varphi _{\alpha \beta}$. Upon substitution into (\ref{V_terms}), the dependence on the $N \! -\! 1$ phases $\phi_{\alpha 1}, \phi_{1 \beta}$ (containing subscript $1$) drops out of $H$.  Then, noting that $\varphi_{1\alpha}=0$, we find that the Borromean coupling involving $\gamma = 1$ reduces to $-g_{ 1 \alpha \beta }\cos \varphi _{\alpha \beta}$, and thereby at low energy enforces $\varphi _{\alpha \beta}=0$, i.e., gapping out $\varphi _{\alpha \beta}$. Hence, modes other than the $N \! -\! 1$ phases $\phi_{1\alpha}$ are gapped, thereby demonstrating $N \! -\! 1$ Goldstone modes in the ordered phase of the SCF$_B$ model (\ref{H})--(\ref{V_terms}).

{\it Dynamic equations and constants of motion.} The convention (\ref{conv}) allows us to cast the $N(N-1)/2$ Hamiltonian equations and their $N(N-1)/2$ complex conjugates into the convenient unified form
\be
i \sigma_{\alpha \beta} \dot{\psi}_{\alpha \beta} \, =\, {\delta H \over \delta \psi_{\beta \alpha} } \, , \qquad 
\sigma_{\alpha \beta} \, =\, \left\{\begin{array}{c}~ 1\, , ~~ \text{if} ~~ \alpha < \beta \, , \\ \!\! -1 \, , ~~\text{if} ~~ \alpha > \beta \, , \end{array}\right. 
\label{EoM}
\ee
where the Hamiltonian $H$ in the r.h.s. of (\ref{EoM}) is written as a functional of the $N(N-1)$ fields $\psi_{\alpha \beta}$ but {\it not} their complex conjugates; we treat all the fields as independent when calculating variational derivatives.

The $N$ elementary symmetry transformations $\hat{U}_\gamma(\varphi)$ generate $N$ Noether's constants of motion and corresponding continuity equations. Given the structure of our symmetry group, only $N-1$ of these $N$ conserved quantities are independent. 

Since $H$ is invariant with respect to transformation $U_\gamma(\varphi)$, we have
\be
{\partial H \over \partial \varphi} = 0 ~~ \Rightarrow ~ \int \! d{\bf r} \! \sum_{\beta (\neq \gamma)}
\left[ {\delta H \over \delta \psi_{\gamma \beta} } {\partial  \psi_{\gamma \beta}  \over \partial \varphi} + 
{\delta H \over \delta \psi_{\beta \gamma} } {\partial  \psi_{ \beta \gamma}  \over \partial  \varphi} \right] = 0. 
\label{rel1}
\ee
Using equations of motion (\ref{EoM}) (to convert variational derivatives of the Hamiltonian into time derivatives of the fields) and the U(1) field transformations (${\partial  \psi_{\gamma \beta}  /\partial \varphi} \, =\, i \psi_{\gamma \beta}$, $ {\partial  \psi_{ \beta \gamma}  / \partial  \varphi} \, =\, - i \psi_{ \beta \gamma}$), we see that (\ref{rel1}) reveals the  additive conserved charges,
\be
Q_\gamma \, = \int   q_\gamma  \, d{\bf r} \, , \qquad q_\gamma  \, = \sum_{\beta \, (\neq \gamma)} \sigma_{\gamma \beta} |\psi_{\gamma \beta} |^2 \, .
\label{Q_alpha}
\ee
Equation (\ref{Q_alpha}) implies the constraint
\be
\sum_{\alpha = 1}^N\, q_\gamma \, =\, 0 \, ,
\label{constr}
\ee
with only $N-1$ independent charge densities $q_\gamma$, as consistent with SCF$_B$'s [U(1)]$^N$/U(1) symmetry and the number of Goldstone modes. 

The corresponding continuity equations are readily found by differentiating the  density $q_\gamma$ of the conserved quantity $Q_\gamma$ with respect to time and using the equations of motion (\ref{EoM}):
\be
\dot{q}_\gamma \, + \, \nabla \cdot {\bf J}_\gamma \, =\, 0 \, , 
\label{cont}
\ee
\be
{\bf J}_\gamma \, =  \sum_{\beta \, (\neq \gamma)} {\bf j}_{\gamma \beta} \, , \qquad  
{\bf j}_{\gamma \beta} \, =\, {i\over 2}\Lambda_{\gamma \beta}  [\psi_{\gamma \beta} \nabla \psi_{ \beta \gamma}  -
\psi_{\beta \gamma}  \nabla \psi_{\gamma \beta} ] \, .
\ee

{\it Ground state.} The ground state is found by minimizing the grand canonical Hamiltonian
\be
H' = H  - \sum_{\alpha=1}^N \lambda_\alpha Q_\alpha = H +
\sum_{ \alpha < \beta} \, (\lambda_\beta - \lambda_\alpha)  |\psi_{\alpha \beta}|^2 ,
\ee
imposing conserved charges $Q_\alpha$ through corresponding Lagrange multipliers $\lambda_\alpha$. 
 The result is the generalized stationary Gross-Pitaevskii equation,  
\be 
{\delta H \over \delta \psi_{\beta \alpha}} \, =\, \sigma_{\alpha \beta} (\lambda_\alpha - \lambda_\beta) \psi_{\alpha \beta} \, .
\label{Stationary}
\ee
Consistent with the linear dependence of densities $q_\alpha$ per \rf{constr}, with only $N-1$ independent ones, the dependence of the ground state on $\lambda_\alpha$'s exhibits one degree of degeneracy, i.e., an invariance under a shift of all $\lambda_\alpha$'s by an $\alpha$-independent constant. 

 The Gross-Pitaevskii equation \rf{Stationary}, with the cubic term (\ref{V_terms}), predicts a non-trivial state with $\psi_{\alpha \beta} \neq 0$. For nonzero and distinct $\lambda_\alpha$'s, Eq.~(\ref{Stationary}) and the equation of motion (\ref{EoM}) predict that the phase $\phi_{\alpha \beta}$ of the  ground-state field $\psi_{\alpha \beta}$
evolves in time according to,
\be
\dot{\phi}_{\alpha \beta} \, =\, \lambda_\beta - \lambda_\alpha \, ,
\label{Phi_dot}
\ee
that is the ground state supports an inter-flavor AC Josephson effect (to be discussed further below), driven by the difference in the corresponding chemical potentials $\lambda_\alpha$.

{\it Hydrodynamics.} With the above analysis and discussion, it is straightforward to see that the dynamical field theory  (\ref{H})--(\ref{V_terms}) at low energy leads to the Borromean hydrodynamics, introduced in Ref.~\cite{BS2024} within the framework of compact-gauge-redundant single-component formalism. 

To this end, we observe, that (by construction), the coupling interaction (\ref{V_terms}) constrains the low-frequency modes to satisfy the vanishing (mod $2\pi$) of triplet plaquette flux condition 
\be
\forall \, \{ \alpha \beta \gamma \}: \quad  \phi_{\alpha \beta} +\phi_{\beta \gamma} + \phi_{\gamma \alpha} \, =\, 2\pi \times \text{integer}
\label{key_cond}
\ee
for the phases of the order parameter $\psi_{\alpha\beta}$, solved by \rf{theta}.
Thus, by construction, the coupling $g_{\alpha\beta\gamma}$  does not couple the phases $\theta_\alpha,\theta_\beta,\theta_\gamma$, implying that the hydrodynamics exhibits local gauge invariance,
\be
\forall \alpha : \quad  \theta_\alpha({\bf r}) \, \to \, \theta_\alpha({\bf r}) \, +\, \phi({\bf r}) \, ,
\label{compact_guage}
\ee
with an arbitrary compact local phase $\phi({\bf r})$, ensuring SCF's insulating characteristic with respect to net charge.  Hence, once expressed in terms of phases $\theta_\alpha$, the currents acquire the standard hydrodynamic form \cite{BS2024}.  

To arrive at the full hydrodynamic Hamiltonian in terms of phases $\theta_\alpha$ and canonically conjugate densities, we consider the Lagrangian formulation of our model. The density of the Lagrangian generating Eqs.~(\ref{EoM}) is given by
\be
{\cal L} \, =\, {i\over 2} \sum_{\alpha \neq \beta} \sigma_{\alpha \beta} \dot{\psi}_{\alpha \beta} \psi_{\beta \alpha} - {\cal H}\, .
\label{L}
\ee
The hydrodynamic action straightforwardly follows from (\ref{L}) utilizing standard long-wave approximations
\[
\dot{\psi}_{\alpha \beta} \, \to \, i(\dot{\theta}_\alpha - \dot{\theta}_\beta) \psi_{\alpha \beta}\, , \quad \nabla \psi_{\alpha \beta} \, \to \, i(\nabla \theta_\alpha - \nabla\theta_\beta) \, \psi_{\alpha \beta} \, ,
\]
\be
{\cal L} \, \to \, - \sum_{\alpha \neq \beta} q_\alpha \dot{\theta}_\alpha - {\cal H} \, ,
\label{hydro}
\ee
where ${\cal H}$ now depends only on the gradients $(\nabla \theta_\alpha - \nabla\theta_\beta)$ and phase-independent densities $|\psi_{\alpha \beta}|^2$, at low energies taken as their ground-state values, set by densities $q_\alpha$. 

The implicit subtlety in the above dynamics is the constraint (\ref{constr}) on charge densities, allowing us to express one of $q_\alpha$'s---say, $q_N$---in terms of the others. Thus, the Lagrangian leads to only $N-1$ independent continuity equations (\ref{cont}), with
\be
{\bf J}_\alpha \, =  \sum_{\beta (\neq \alpha)}  \Lambda_{\alpha \beta}  \, |\psi_{\alpha \beta}|^2 \, (\nabla \theta_\alpha - \nabla\theta_\beta) 
\label{J_hydro}
\ee
and $N-1$ conjugate equations,
\be
\dot{\theta}_\alpha - \dot{\theta}_N \, =\, - 
{\partial {\cal H} \over \partial q_\alpha} \, .
\ee
Utilizing the U(1) gauge redundancy of the model, allows us to set $\theta_N \equiv 0$. This leads to $N\! -\! 1$ canonically conjugate pairs $(\theta_\alpha, q_\alpha)$ with the Hamiltonian density ${\cal H} \equiv {\cal H}(\{ \nabla\theta_\alpha, q_\alpha\})$, $\alpha =1,\, 2, \ldots , (N\!- \!1)$.

Alternatively, the gauge invariance allows us to formally treat all the $N$ densities $q_\alpha$ on equal footing as independent dynamical variables, with the constraint (\ref{constr}) emerging {\it dynamically} (cf.~\cite{BS2024}). Now 
we generalize the definition of the quantities $|\psi_{\alpha \beta}|^2$---and thus the function ${\cal H}$---by requiring that these are certain (not uniquely defined) functions of {\it all} the $N$ variables $q_\alpha$ provided the values of the functions coincide with the original definitions under the constraint (\ref{constr}). Now the theory produces all $N$ continuity equations (\ref{cont})---with $N$ currents (\ref{J_hydro})---and $N$ equations of the Beliaev--Josephson--Anderson type
\be
\dot{\theta}_\alpha \, =\, - 
{\partial {\cal H} \over \partial q_\alpha} \, .
\label{BJA}
\ee
This is a Hamiltonian theory with compact-gauge invariant Hamiltonian density ${\cal H} \equiv {\cal H}(\{ \nabla\theta_\alpha, q_\alpha\})$, $\alpha =1,\, 2, \ldots , N$.
The sum of all the continuity equations now yields the relation (Noether's local constant of motion enforced by the compact-gauge symmetry, cf.~\cite{BS2024}) 
\be
{d\over dt}\sum_{\alpha =1}^N \, q_\alpha \, =\, 0
\label{constr2}
\ee
meaning that the constraint (\ref{constr}) is consistent with (being preserved by) the equations of motion.

{\it Counterflow AC Josephson effect.}  The AC Josephson effect is the hallmark of superfluidity, distinguishing it from other class of states that break O(2) symmetry. Its counterflow counterpart naturally arises in super-counterfluids. Moreover, AC Josephson effect is where the formulation in terms of fundamental tensor fields $\psi_{\alpha \beta}$ clarifies the physics, each component associated with a distinct counter-flow Josephson tunneling channel. The simplest and conceptually instructive case is one involving just one channel, say, $\psi_{\alpha_0 \beta_0}$, realized by a weak link ($\alpha_0\beta_0$) between two super-counterfluids. The counterflow Josephson oscillations are then characterized by a single frequency
\be
\omega_{\alpha_0 \beta_0} \, =\, | \Delta \dot{\phi}_{\alpha_0 \beta_0}| =
| \Delta \mu_{\beta_0} - \Delta \mu_{\alpha_0}|  \, ,
\ee
where $\Delta \mu_{\beta_0}$ and  $\Delta \mu_{\alpha_0}$ are the differences of corresponding chemical potentials.

{\it Vicinity of the} [U(1)]$^N$/U(1) $\to$ [U(1)]$^N$ {\it transition. }
The dynamical field theory of SCF$_B$ discussed above captures low-energy properties of $N$-component bosons on a lattice in the Mott regime with respect to the net current. The Mott ground state can only be formed if the net filling factor is integer and the interactions sufficiently exceed kinetic energy.   This leads to a natural question of SCF$_B$-SF$_N$ phase transition to an $N$-component superfluid that breaks  full [U(1)]$^N$ symmetry. It can be driven by slightly doping away from commensurate filling or weakening the interaction. Corresponding physics is naturally captured by the ``doped" version of dynamical model containing a single-component sector represented by fields $\psi_\nu$, $\nu = 1, \, 2, \, \ldots , \, N$ transforming under [U(1)]$^N$ as its basic representations. 

The dopped model includes a coupling of the single-component fields $\psi_\nu$ to the tensor fields $\psi_{\alpha \beta}$, that minimally is given by
\be
{\cal H}_\text{couple}\, \to \, {\cal H}_\text{couple}\, - \, \sum_{\alpha < \beta} \eta_{\alpha \beta} (\psi_{\alpha \beta} \psi_\alpha^* \psi_\beta \, +\, \psi_{ \beta \alpha} \psi_\alpha \psi_\beta^*) \, , 
\label{doppedH}
\ee
where $ \eta_{\alpha \beta}$ is a coupling constant.

The symmetry transformation $\hat{U}_\alpha (\varphi)$ is now augmented with $\psi_\alpha \to \psi_\alpha e^{i\varphi}$, which renders all $N$ transformations independent. Corresponding $N$ Noether's constants of motion are also independent.
Their explicit form is readily obtained by a straightforward generalization of the above-discussed procedure. We get $N$ independent continuity equations of the form (\ref{cont}) with the upgraded expressions for the densities $q_\alpha$ and the currents ${\bf J}_\alpha$:
\be
q_\alpha \, \to \, q_\alpha  \, +\,  |\psi_\alpha|^2 \, , \qquad {\bf J}_\alpha  \, \to \, {\bf J}_\alpha \, + \, {\bf j}_\alpha \, ,
\ee 
where ${\bf j}_\alpha$ is a standard single-component current of the Gross-Pitaevskii equation for the component $\alpha$. Summing all the continuity equations
yields the continuity equation for the net-current mode:
\be
\dot{q} + \nabla \cdot {\bf j} = 0\, , \qquad q = \sum_{\alpha=1}^N |\psi_\alpha|^2 \, , \qquad {\bf j} = \sum_{\alpha=1}^N \, {\bf j}_\alpha\, .
\ee

{\it First-order phase transitions.}  The presence of cubic terms in the Borromean super-counterflow model (\ref{H})--(\ref{V_terms}) and its doped generalization \rf{doppedH} allows for a generic mechanism of a first-order finite-temperature and quantum insulator (normal gas) to SCF$_B$ phase transition.  In both the genuine and weakly doped Borromean system, such a regime happens naturally if the system features exact or approximate permutation symmetry between the fields $\psi_{\alpha \beta}$.  A weak doping is irrelevant to the nature of the normal-to-Borromean phase transition because the condensation of the weak flavor-neutral mode takes place only at a much lower temperature. We note in passing that if the doping is not weak, then the first-order transitions can happen under conditions of fine-tuning the fields $\psi_\alpha$, $\psi_\beta$,  and $\psi_{\alpha \beta}$ towards almost  simultaneous condensation, as was discussed in Ref.~\cite{Kuklov2004}; see also Ref.~\cite{Radzihovsky2004} for a single-component analog.

{\it Gauged version of the doped model.}
Our SCF$_B$ model can be generalized to the case where dopped fields $\psi_\mu$ are charged and interact with the common $U(1)$ gauge field, namely to a multicomponent superconductor. In the case of a particular physical interest of a multi-band superconductor, weak composite-symmetry-breaking inter-band hybridization 
\be
\sim(\psi_{\alpha \beta} + \psi_{\beta \alpha}) \qquad \quad 
\label{Josephson}
\ee
will generically appear. In the case when these Josephson couplings are frustrating, i.e.,  phases of the fields $\psi_{\alpha \beta}$ minimizing  (\ref{Josephson}) are incompatible with the (dominant) ${\cal H}_\text{couple}$  in (\ref{V_terms}) and therefore cannot be equal to 0 or $\pi$, such superconductor exhibits a broken time-reversal symmetry \cite{Bojesen2013,Bojesen2014}.

In the gauged case, it is appropriate to consider both weak and strong doping as well as weak and strong coupling to the gauge field. This is because now the flavor-neutral net-current mode is guaranteed to be distinctively different from the other modes due to the Anderson-Higgs effect enforcing the  [U(1)]$^N$/U(1) symmetry of the charge-neutral modes. As with the weakly doped electrically neutral model, finite temperature can suppress the superconducting order leaving behind super-counterfluid state \cite{Babaev2004} (or the most peculiar normal state with broken time-reversal symmetry in the case with frustrating Josephson coupling \cite{Bojesen2013,Bojesen2014}).

{\it Concluding remarks.}  With respect to the universal long-wave dynamical and statistical properties, the composite counterflow (net-charge-neutral) orders are not distinguishable from single-component ones and should be treated accordingly. The details of the microscopic origin of the counterflow modes are not only irrelevant but can even prove misleading. In the low-energy context, the counterflow order parameters emerge as fundamental field-theoretical representations of the composite symmetry group [U(1)]$^N$/U(1). 

Developing this concept, we introduced a representation of the  [U(1)]$^N$/U(1) symmetry group in terms of the  inter-flavor complex-valued tensor field (with a cubic phase-locking to project down to $N-1$ Goldstone modes) as a model of super-counterfluid as well as  closely related systems such as doped counterflow superfluids and Borromean insulators. 
 
Within the paradigm of universality of the long-energy properties---being controlled exclusively by corresponding symmetry group---approach presented here provides a solid basis for a microscopic derivation of the universal compact-gauge invariant Borromean hydrodynamics, classification of topological defects/counterflow supercurrent states, revealing the counterflow Josephson effect, as well as addressing the critical properties of Borromean (and closely related) systems in terms of corresponding order parameters. 

Our SCF$_B$ field theory is naturally promoted to an effective field theory describing Borromean criticality consistent with the expectation (E. Babaev, private communication) that the effective action should feature cubic terms enforcing the first-order character of the transition in dimensions larger than two. 
 
 \begin{acknowledgements} 
 The authors are grateful to Egor Babaev for stimulating discussions. They thank NORDITA for hospitality and support during the workshop Beyond Standard Superconducting and Superfluid States, where this work was initiated. AK and BS acknowledge support from the National Science Foundation  under Grants  DMR-2335905 and DMR-2335904. LR acknowledges support by the Simons Investigator Award from The James Simons Foundation. 
\end{acknowledgements}

\bibliography{Borromean_FT}

\end{document}